\documentstyle[aps,prl,twocolumn,epsf]{revtex}
\begin{document}
\epsfverbosetrue
\title{The critical amplitude ratio of the susceptibility
in the random-site two-dimensional Ising model}
\author{Lev N. Shchur \dag\ and Oleg A. Vasilyev}
\address{Landau Institute for Theoretical Physics, 142432 Chernogolovka,
Russia \\
\dag\  \sl e-mail: lev@itp.ac.ru}
\maketitle
\begin{abstract}

We present a new way of probing the universality class of the
site-diluted two-dimensional Ising model. We analyse Monte Carlo data
for the magnetic susceptibility, introducing a new fitting procedure in
the critical region applicable even for a single sample with quenched
disorder. This gives us the possibility to fit simultaneously the
critical exponent, the critical amplitude and the sample dependent
pseudo-critical temperature.  The critical amplitude ratio of the
magnetic susceptibility is seen to be independent of the concentration
$q$ of the empty sites for all investigated values of $q\le 0.25$. At
the same time the average effective exponent $\gamma_{eff}$ is found to
vary with the concentration $q$, which may be argued to be due to
logarithmic corrections to the power law of the pure system. This
corrections are canceled in the susceptibility amplitude ratio as
predicted by theory. The central charge of the corresponding field
theory was computed and compared well with the theoretical predictions.

\end{abstract}


\draft

The effect of impurities on the critical behavior is one of the central
questions of phase transition phenomena~\cite{DM-rev,our-rev}.  According to
the Harris criterion, weak concentrations of impurities do not affect the
critical exponents if $d\nu>2$, where $d$ is the system dimension, and $\nu$
is the exponent of correlation length in the pure system \cite{Harris74}.
In the two-dimensional Ising model (2DIM) one has $d\nu=2$. In other words,
the 2DIM is a marginal case, where logarithmic corrections may become
important in the vicinity of the phase transition, while critical exponents
are not changed, as has been found analytically for the specific heat and
correlation length \cite{DD} as well as for the magnetic susceptibility
\cite{correct}. Indeed, the widely accepted picture is that {\em weak bond}
dilution changes the critical behavior of the correlation length of the pure
model $\xi\simeq 1/|\tau|$ by a logarithmic term into
$$
\xi\simeq\frac{\sqrt{1+\frac{4}{\pi}g_0\;\ln\;\frac{1}{|\tau|}}}{|\tau|},
$$
where $\tau=(T-T_c)/T$ is the reduced temperature, with $T_c$ being the
critical temperature and $g_0$ is a coefficient proportional to the strength
of disorder (strictly speaking, $g_0$ is the central charge of the $N=0$
Gross-Neveu model related to the 2DIM, see \cite{DD}-\cite{Plechko}, and
$g_0=0$ for the pure 2DIM).  Similarly, the critical exponent of the
magnetization $M$ and the magnetic susceptibility $\chi$ exponents remains
the same as for the pure 2DIM, with the critical behavior being modified by
logarithmic factors. This prediction for weak bond disorder was confirmed
numerically in a number of papers \cite{our-rev,Wang,RAJ}.

In the following, we study the universality class of the {\em site-diluted}
Ising model. The phase diagram of this model contains two crucial points,
the pure Ising fixed point at zero concentration of impurities, $q=0$, and
the percolation fixed point at $q=q_c=0.407254$~\cite{Ziff}. The
site-diluted 2DIM was investigated numerically by quite a few
authors~\cite{Kim}-\cite{SSV} in the last few years. The interpretations of
the results are rather controversial. Some of authors
claim~\cite{Kim,change} that the critical exponents, e.g. that of the
magnetic susceptibility, $\gamma$, vary with the impurity concentration $q$,
but that the ratio, say, $\gamma/\nu$ (where $\nu$ is the critical exponent
of the correlation length) does not change. Others~\cite{remain,SSV}
concluded that impurities lead, for the susceptibility as well as for other
quantities, only to logarithmic corrections as for weak bond dilution. In
all cases, the analyses have been performed either above, at or below the
critical point, $T_c$.

The main aim of the present study is to identify the universality class by
computing the critical amplitude ratio of the magnetic susceptibility
$\chi$, thereby comparing data both below and above $T_c$.

It is well known that the universality class is not only characterized by
its critical exponents but also by the critical amplitude
ratios~\cite{PHA-rev}. For instance, in zero external magnetic field the
critical behavior of the magnetic susceptibility $\chi$ is given by
$\chi\simeq\Gamma\;\tau^{-\gamma}$ in the symmetric phase and by
$\chi\simeq\Gamma'\;|\tau|^{-\gamma'}$ in the ordered phase, with $\Gamma$
and $\Gamma'$ being the critical amplitudes and $\gamma$ and $\gamma'$ are
the critical exponents. $\Gamma/\Gamma'$ is the critical amplitude ratio.

Note that the critical amplitude ratio is very often quite sensitive to the
universality class. The basic idea is that if the values of exponents vary
from the pure Ising model to the percolation ones as reported
in~\cite{Kim,change}, then one could expect a variation of the critical
ratio $\Gamma/\Gamma'$ as well. So, we may expect that the critical ratio
will change from the value $\Gamma/\Gamma'=37.69365$ known for the pure
two--dimensional Ising model~\cite{WMTB} to the percolation limit value
$\Gamma/\Gamma'\approx 170$~\cite{Ziff-pc}. Clearly, such a variation of
about four times is much easier to check than the variation of the exponents
which is reported to be approximately 10 percent.

To investigate the critical behavior of the magnetic susceptibility and
the corresponding amplitude ratio, we carried out extensive simulations
of the 2DIM with site dilution. Numerical calculations of the critical
amplitudes of the susceptibility are known to be difficult, even for the
pure Ising and Potts models~\cite{DC,DBC,CTV,SalSok,DSS} due to
finite-size effects and analytical corrections to scaling. For the
diluted case, the problem becomes even more delicate because of a lack
of quantitative knowledge on the critical temperature and the possible
crossover associated with the randomness induced logarithmic factors.

The crucial technical idea of the present work is a heuristic fitting
procedure to determine the critical amplitude ratio.

The paper is organized as follows. First, we will summarize the main
results. Then, we will describe the fitting procedure and demonstrate how it
works for the pure model. Then, specific results will be presented and
discussed.

The main results of our study can be summarized as follows:

1. {\em New fitting procedure:} A sample dependent pseudo-critical
temperature $T_c^*(q;L,i)$ ($i$ refers to the sample and $L$ is the linear
dimension of the square lattice) follows from fitting the susceptibility in
the high and low temperature phases close to criticality, to
$\chi(\tilde{\tau})\simeq \Gamma\;\tilde{\tau}^{-\gamma_{eff}}$ and
$\chi(\tilde{\tau})\simeq \Gamma'\;|\tilde{\tau}|^{-\gamma'_{eff}}$,
respectively. $\gamma_{eff}, \gamma'_{eff}$ are average effective critical
exponents determined in a range of temperatures, where finite--size effects
and corrections to scaling may be neglected;
$\tilde{\tau}=|T-T_c^*(q;L,i)|/T$ is the sample dependent reduced
temperature relative to $T_c^*$. The fitting condition is the equality of
the effective critical exponents $\gamma_{eff}$ and $\gamma'_{eff}$, which
is applicable if the corrections to scaling are not large in the critical
region as is the case for the 2D Ising model.

2. {\em Universality of the critical amplitude ratio for the magnetic
susceptibility:} We estimated the ratio $\Gamma/\Gamma'$ by fitting our
Monte Carlo data for lattices with linear dimension $L= 256$. The ratio
seems to remain constant for all concentrations of impurities we
considered, i.e. for $q$ in the interval $[0,0.25]$. This behavior may
be interpreted as a manifestation of the universality class of the
site--dilute Ising model being independent of the degree of dilution. It
should be noted that theory predicts the cancellation of the logarithmic
corrections~\cite{DD,DC,DBC} in the ratio of high-temperature and
low-temperature susceptibilities and, thus, predicts the universality of
critical amplitude ratio, at least for small concentration of impurities
where DD theory \cite{DD} are applicable.

3. {\em Weak and strong dilution regions:} We found the value of
dilution $q^*=0.1$ up to which the predictions of the weak-dilution
theory~\cite{DD,correct,Plechko} works well could be explained as
concentration at which two intrinsic lengths coincide. The first is the
average distance between impurities~\cite{DD} $l_i\propto exp(1/g)$. The
second is the percolation correlation length $\xi_p\propto (q_c-q)^{-4/3}$
characterizing the size of the holes formed on the lattice by the diluted
sites. For the concentration of diluted sites larger than $q^*$ the theory
of weak dilution cannot be applied because the size of the conglomerates
of the diluted sites becomes larger than the average distance between the
diluted sites and the effective interaction between impurities should be
taken into account.

4. {\em Variation of the average effective exponent $\gamma_{eff}$:} We
find that the average effective critical exponent $\gamma_{eff}$ varies with
the concentration of impurities, similarly to previous observations
~\cite{Kim,change,remain}. The variation may be attributed to logarithmic
corrections, growing linearly with the small concentration of dilution up to
the concentration $q^*=0.1$.

5. {\em Central charge $g_0$ of the $N=0$ Gross-Neveu model:} We extract
from numerics its dependence upon the concentration of diluted sites and
found the values coincide well with ones predicted by exact expression for
the charge in terms of the site-diluted model~\cite{Plechko} for the
dilution concentration $q$ up to about $q^*=0.1$.

{\sl The model:} Each site of a square lattice is either occupied by an
Ising spin, $S_i = \pm 1$ or not. The fraction of empty sites is denoted by
$q$. The positions of non--occupied sites were generated using the
shift-register generator with lags $(9689,417)$ which is known to be
appropriate for selecting randomly lattice sites ~\cite{SHB}. At each
concentration of empty sites $q$, chosen to be
$q=0,0.03,0.07,0.1,0.15,0.18,0.2,0.22,$ and 0.25, the number of realizations
(or samples) in the simulations ranged between ten and twenty five.

It seems obvious that the singular part of the magnetic susceptibility stems
from fluctuations of spins belonging to the percolating cluster of occupied
sites~\cite{DM-rev,SA-rev}. Those spins which are disconnected from the
percolation cluster do not contribute to the critical behavior. Accordingly,
we took into account only the fluctuations of the spins in the largest
cluster, thereby reducing the ``noise'' in the susceptibility due to the
small clusters (and reducing the computation time).

{\sl The simulations:}  For each concentration of empty sites, we computed
the magnetic susceptibility $\chi(T,q;L,i)$ in the critical region at about
forty temperatures. The temperatures were chosen in the interval
$\tau_r<|\tau|<\tau_a$ where $\tau_r$ is the rounding temperature
$\tau_r\simeq 1/L$ \cite{FiFe}, above which finite-size effects may be
neglected, and $\tau_a$ is the reduced temperature, above which corrections
to scaling become important \cite{our-rev}. For the pure Ising model,
$\tau_a$ may be estimated from the exact solution~\cite{WMTB}
\begin{equation}
k_BT\;\chi(\tau)=\Gamma\;
|\tau|^{-\gamma}\;(1+e_\chi\tau+...)
\label{exact}
\end{equation}
with $e_\chi=0.07790315$; the corrections to scaling, $e_\chi\tau$, become
important only at rather large reduced temperatures, say,
$\tau>\tau_a\approx 0.13$. Presumably, dilution does not affect $\tau_a$
significantly \cite{our-rev}.

In the simulations, we used the one-cluster-flip algorithm ~\cite{Wolff},
discarding the first $10^4$ clusters for thermal equilibration. Totally,
$10^5$ clusters were generated for each sample $i$, at given values of $q$
and $T$.

{\sl The fitting procedure:} The fitting procedure is based on the
assumption that the critical exponent of the magnetic susceptibility
$\gamma$ takes the same value below and above the critical temperature.
Thence, one may determine the sample dependent pseudo-critical temperature
$T_c^*(q;L,i)$, as described above, with the average effective critical
exponents, $\gamma_{eff}$ and $\gamma_{eff}'$ coinciding in the interval
$[\tau_r, \tau_a]$. In this way, we extracted sample dependent critical
amplitudes $\Gamma(i)$ and $\Gamma'(i)$ as well as the effective critical
exponent $\gamma_{eff}(i)$. 

{\sl Assessment of the accuracy of our approach:} To check the accuracy of
our approach, we studied also the pure Ising model, where the critical
amplitudes are known exactly.

\begin{figure}
\epsfxsize=\columnwidth
\epsfysize=\columnwidth
\epsfbox{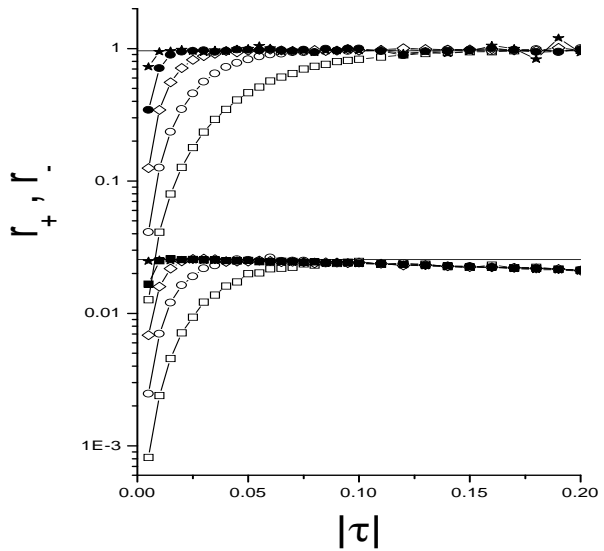}
\caption{The ratios of the computed
magnetic susceptibility divided by the singular part including
the leading corrections to scaling (exactly known (\ref{exact})
$r=|\tau|^{-1.75} (1+e_{\chi})$)
for the pure Ising model in high ($r_+$) and low ($r_-$) temperature phases
for system sizes $L=$ 16 (open boxes), 32 (open circles), 64 (open diamonds),
128 (closed boxes), 256 (closed stars). See text for the futher details.}
\label{xpxm}
\end{figure}

First, we considered the ratio of $\chi(\tau)=\langle
M^2\rangle-\left(\langle |M|\rangle\right)^2$, as computed in the
simulations, to the singular part of the magnetic susceptibility
multiplied by the leading correction to scaling (i.e.
$\chi(\tau;L)/\tau^{-\gamma}\;(1+e_\chi\tau)$, see Eq. (1)), with the
linear dimension $L$ of the square lattice ranging from 16 to 256. To
reduce finite-size effects, we set in the simulations $\langle
|M|\rangle=0$ in the symmetric, high temperature phase, $T_c>0$. The
results are shown in Fig~\ref{xpxm}, where the upper solid line
corresponds to the exact value of critical amplitude above the critical
point, $\Gamma=0.962582$, and the lower solid line corresponds to
$\Gamma'=0.025537$. Clearly, the critical amplitudes are approximated
closely in the range $\tau_r<|\tau|<\tau_a$, especially for $L$= 256. Note
some deviations of the magnetic susceptibility from the asymptotically
exact value deeply in the ordered phase, which could be eliminated taking
into account the background term $D_0=-0.104133...$~\cite{KAP}. Actually,
we have checked the sensitivity of the fitting results to the inclusion of
this term. One might estimate the accuracy of the numerical value of the
critical amplitude ratio $\Gamma/\Gamma'$ (=37.69..) extracted from the
Monte Carlo data to be about $5$ percent.

\begin{figure}
\epsfxsize=\columnwidth
\epsfysize=\columnwidth
\epsfbox{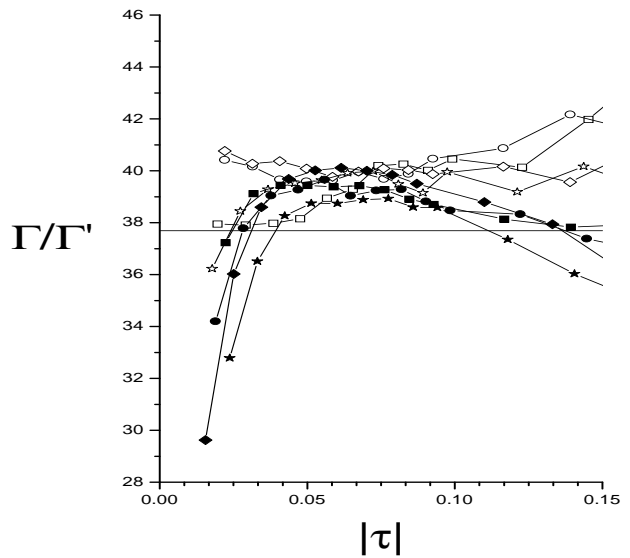}
\caption{
The critical amplitude ratio computed as the ratio of susceptibility data
in the ordered and the symmetric phases $\chi(\tau)/\chi(-\tau)$ for various
concentrations of dilution, $q$=0 (open boxes), 0.03 (open circles),
0.07 (open diamonds), 0.10 (open stars), 0.15 (closed boxes),
0.18 (closed circles), 0.20 (closed diamonds) and 0.25 (closed (stars)
for the system size $L=256$.}
\label{ratio}
\end{figure}

Next, with $L=256$ and for the different concentrations of dilution $q$,
we consider the ratio of the Monte Carlo data for the magnetic
susceptibility $\chi/\chi'$, averaged over the various samples. The
temperature scale has been determined by using $T_c^*(q)$. Obviously, as
shown in Fig~\ref{ratio}, the ratio is (nearly) constant in the range
$[\tau_r,\tau_a]$. In that temperature range, one may expect that both
finite-size effects and corrections to scalings are negligible. However,
crossover terms in the presumed logarithmic corrections are expected to
depend rather sensitively on the degree of dilution, $q$. However, if
they are of similar nature above and below the critical point, our
approach circumvents this difficulty, as seems to be consistent with the
results depicted in that figure. It is a commonly accepted
picture~\cite{DC,DBC} that the logarithmic corrections should canceled
in the ratio of the high-temperature and low-temperature
susceptibilities. Our results clearly support such scenario.

{\sl Specific results and discussion:} Our results on the critical
amplitude ratio and the pseudocritical temperature, averaging over the
sample-dependent results, for $L$=256 and various concentrations of
dilution, are presented in Table~\ref{res} with $1$-$\sigma$ error bars
in parentheses.  One may emphasize that the pseudocritical temperature
computed by us differs from the values of the ``true'' critical
temperature reported by other authors~\cite{Kim,change} (usually based
on finite-size analyses) by less than $O(\frac1L)$, reassuring us of the
correctness of our approach. It is remarkable, that the ratio of
critical amplitudes remains constant within the error bars, while the
critical amplitudes by themselves grow by almost a factor of three, when
varying the dilution from $q$= 0 to 0.25. We limit our simulations at
concentration $q=0.25$ because larger dilutions need larger system sizes
in order to have a reasonably wide critical region. Unfortunately,
simulations of larger sizes is out of computation power at the moment.
The (small) deviation of the critical amplitude ratio in the pure case
from the exact value could be reduced by including in the fit the
background term $D_0$. The fit in the reduced pseudo-critical
temperature window $0.015<\tilde{\tau} <0.2$ gives $D_0=-0.07(2)$ and
ratio $\Gamma/\Gamma'=38.7(4)$ deviates only by two-sigma from the exact
value.

\begin{figure}
\epsfxsize=\columnwidth
\epsfysize=\columnwidth
\epsfbox{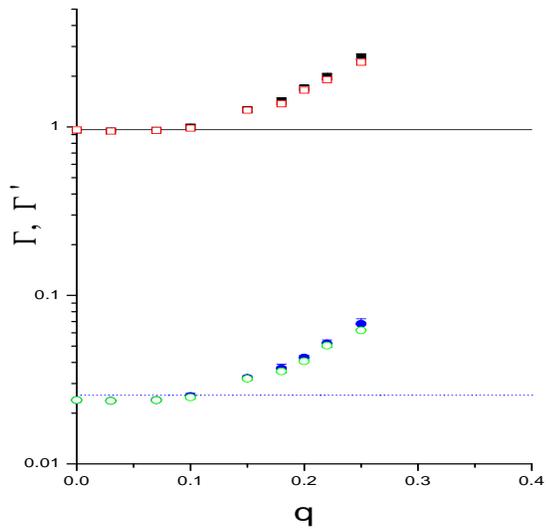}
\caption{The critical amplitudes $\Gamma$ (closed boxes)
and $\Gamma'$ (closed circles) of the magnetic
susceptibility as function of the concentration of empty sites $q$.
Open signs denote results of the fit to the magnetic susceptibility first
averaged over samples.
Horizontal lines refer to the exact values for the pure Ising
model.}
\label{ampl}

\end{figure}

Figure~\ref{ampl} shows the dilution--dependent critical amplitudes $\Gamma$
and $\Gamma'$, given in Table~\ref{res}, as well as the values obtained from
the fit to the magnetic susceptibility averaged first over samples. The nice
agreement of the results shows that the order of the averaging plays no
role. The statistical errors are slightly lower in the former case, as
already mentioned by Wiseman and Domany~\cite{WD}.

\begin{figure}
\epsfxsize=\columnwidth
\epsfysize=\columnwidth
\epsfbox{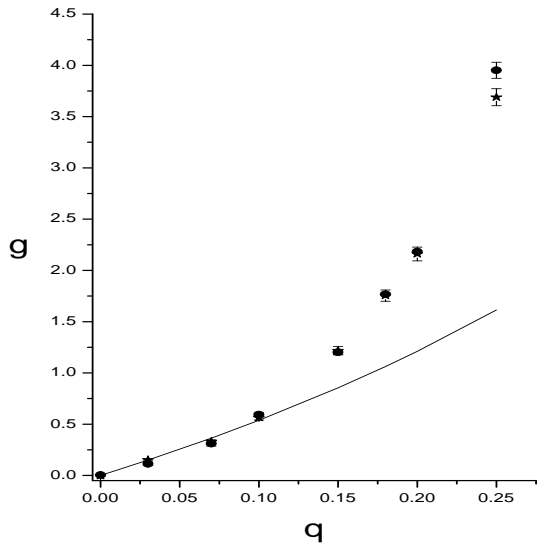}
\caption{The coefficient $g=4\/\pi g_0$  as a function of the impurity
concentration. The solid line denotes the analytic result of Ref.~8. Circles
(stars) correspond to fits of Monte Carlo data for the susceptibility
in the low (high) temperature phase.}
\label{plechko}
\end{figure}

The effective exponent $\gamma_{eff}$ is also given in the Table~\ref{res}.
Obviously, starting in the pure case, $\gamma_{eff}$ first increases quite
rapidly with dilution, but changes only mildly at stronger dilution, $q
\approx 0.2$. The initial variation of $\gamma_{eff}$, at weak dilution, may
be explained quantitatively. Analytically, the magnetic susceptibility has
been calculated \cite{ADD,Plechko} to have the form
\begin{equation}
\chi(\tau)=\Gamma |\tau|^{-7/4} (1+0.07790315 \tau) \left(1-g\;\; \ln \;
|\tau|\right)^{7/8} 
\label{full} 
\end{equation} 
with the coefficient $g$
given~\cite{Diss} by
\begin{equation}
g=\frac{4}{\pi}g_0=\frac{4}{\pi}\frac{8}{(1+\sqrt{2}/\pi)^2}\frac{q}{1-q}
\approx 4.843 \frac{q}{1-q}. 
\label{coeff} 
\end{equation}
In Fig~\ref{plechko}, the coefficient $g$, from (\ref{coeff}), is plotted
together with the fit of the magnetic susceptibility data, below and above
the critical point, to (\ref{full}). The results are expected to agree for
weak dilution; indeed, the agreement holds up to $q$=0.1. Otherwise,
pronounced deviations set in for stronger dilution. In fact,
Fig~\ref{plechko} demonstrates two facts. On one hand, it is a check of
the recent analytic, supposedly exact result by Plechko on the coefficient
$g$. On the other hand, it is consistent with (\ref{full}), which, in
turn, readily explains the variation of the average effective exponent
$\gamma_{eff}$, computed in the interval $[\tau_r,\tau_a]$, with dilution
(see also Ref. \cite{Wang}). Taking the logarithmic derivative of
(\ref{full}) and properly choosing the temperature interval for the
averaging~\cite{BER} of the $ln\; \tau$ term, one can get a linear
dependence of the effective exponent $\gamma_{eff}\propto 7/4\; (1+a q)$
with a coefficient $a$ of the order of unity. This coincides quite well
with the values of $\gamma_{eff}$ in the Table for $q<q^*\approx 0.1$.

It seems that this value $q^*\approx 0.1$ could be explained at follows.
There are two lenghts for the diluted model in addition to the two length
scales one has in the Ising model: namely the correlation length
$\xi\propto 1/\tau$ and the system size $L$. The first length $l_i\propto
exp(-1/g)$ is defined~\cite{DD} by the value at which the term $g\;\; \ln
\; |\tau|$ in (\ref{full}) becomes of the order of unity. The next length
is the percolation correlation length~\cite{SA-rev} $\xi_p\propto
(q_c-q)^{-4/3}$. One could check that these two lenghts coincide for
$q\approx 0.1$. Physically this means that for $q>q^*$ the disorder is no
weaker than assumed in the theories mentioned above.

The critical amplitudes are practically the same for the weak dilution
$q<q^*$. Also the effective exponents are visibly modified in this region
by logarithmic corrections. They start to grow in the region of the
``strong" dilution $q^*<q<q_c$ and their ratio seems to remain unchanged.
Probably, one could expect to see a crossover regime from Ising to
percolation universality class in the very vicinity of the percolation
point.

{\sl Conclusions:} Of course, reliable Monte Carlo data on even larger
system sizes may still be desirable in order to check the universality
of the ratio of critical amplitudes for the magnetic susceptibility, as
suggested by our study. One should also try to analyse systems with even
stronger dilution, i.e. closer to the percolation limit~\cite{SA-rev}.
Nevertheless, our data already provide evidence that the two-dimensional
Ising model with site-dilution is described by the same critical
exponent (modulo logarithmic corrections) and the same critical
amplitude ratio of the magnetic susceptibility as the pure Ising model,
in the range of dilutions investigated. The critical amplitude ratio
$\Gamma/\Gamma'$ is always quite close to the pure Ising value $37.69$
and far away from the percolation value $160-200$ (which is known from
simulations~\cite{Ziff-pc,DBC}). The small apparent variation of the
average effective exponent with the degree of dilution may be explained
as being due to logarithmic corrections.

\acknowledgments

We acknowledge useful discussions with B. Berche, P. Butera, B. Derrida, W.
Janke and J.-K. Kim. Our special thanks to W. Selke who asked many
interesting questions we tried to answer in the present paper, and for
his useful comments to this paper. The work has been supported by grants
from NWO, INTAS and RFBR. O.A.V. thanks the Landau stipendium committee
(Forschungzentrum/KFA J\"ulich) for support.

\begin{table}
\caption{Results of the fit to magnetic susceptibility for the lattice
size $L=256$.}
\center
\begin{tabular}{|l|l|l|l|l|l|}
$q$ & $\Gamma'$ & $\Gamma$& $\gamma_{eff}$ & $T_c$   & $\Gamma/\Gamma'$ \\
\hline\hline
exact& 0.02553 & 0.96258 & 1.75 & 2.26918 & 37.685 \\ \hline
0.00& 0.0239(8) & 0.960(5)& 1.757(1) & 2.2684(7) & 40.21(20)  \\ \hline
0.03& 0.0237(2) & 0.946(8)& 1.814(3) & 2.1610(2) & 39.96(19)\\ \hline
0.07& 0.0239(4) & 0.96(2) & 1.885(6) & 2.0146(3) & 39.98(38)\\ \hline
0.10& 0.0253(9) & 1.00(3) & 1.93(1)  & 1.9032(5) & 39.65(59)\\ \hline
0.15& 0.0326(1) & 1.27(3) & 1.95(1)  & 1.7098(4) & 39.12(55)\\ \hline
0.18& 0.037(2)  & 1.42(6) & 1.99(2)  & 1.5905(7) & 38.92(85)\\ \hline
0.20& 0.043(1)  & 1.70(5) & 1.97(1)  & 1.5103(4) & 39.98(58)\\ \hline
0.22& 0.052(3)  & 1.99(8) & 1.96(2)  & 1.426(1)  & 39.1(1.4)\\ \hline
0.25& 0.066(4)  & 2.57(9) & 1.97(2)  & 1.298(1)  & 40.7(1.9)
\end{tabular}
\label{res}
\end{table} 

\end{document}